\documentclass[12pt]{article}
\usepackage{epsfig}
\usepackage{amsfonts}
\usepackage{amsmath}
\usepackage{amssymb}
\usepackage{graphicx}

\usepackage{color} 
  \newcommand{\beq}{\begin{equation}}
  \newcommand{\eeq}{\end{equation}} 
  \def\nuc#1#2{\relax\ifmmode{}^{#1}{\protect\text{#2}}\else${}^{#1}$#2\fi}
  \def\itnuc#1#2{\setbox\@tempboxa=\hbox{\scriptsize\it #1}
    \def\@tempa{{}^{\box\@tempboxa}\!\protect\text{\it #2}}\relax
    \ifmmode \@tempa \else $\@tempa$\fi}
  
\newcommand{\simge}{\hspace*{0.2em}\raisebox{0.5ex}{$>$}
     \hspace{-0.8em}\raisebox{-0.3em}{$\sim$}\hspace*{0.2em}}
\newcommand{\simle}{\hspace*{0.2em}\raisebox{0.5ex}{$<$}
     \hspace{-0.8em}\raisebox{-0.3em}{$\sim$}\hspace*{0.2em}}

\newcommand{\dslash}[1]{#1 \llap{/\kern-0.5pt}}
\newcommand{\Dslash}[1]{#1 \llap{/\kern+1.2pt}}
\newcommand{\DDslash}[1]{#1 \llap{/\kern+2.3pt}}
\newcommand{\dslashh}[1]{#1 \llap{/\kern+1pt}}

\def\Lag{\mathcal{L}}

\def\bdm{\begin{displaymath}}
\def\edm{\end{displaymath}}

\begin{document}

\begin{titlepage}

\vspace*{1.5cm}

\begin{center}
{\Large\bf Weinberg's Compositeness}
\\

\vspace{2.0cm}

{\large \bf U. van Kolck}

\vspace{0.5cm}
{
{\it Universit\'e Paris-Saclay, CNRS/IN2P3, IJCLab,
\\
91405 Orsay, France}
\\
and
\\
{\it Department of Physics, University of Arizona,
\\
Tucson, AZ 85721, USA}
}

\vspace{1cm}

\today

\end{center}

\vspace{1.5cm}

\begin{abstract}
Nearly 60 years ago Weinberg suggested a criterion for
particle ``compositeness'', which has acquired new life
with the discovery of new, exotic hadrons.
His idea resonates with model-based intuition.
I discuss the role it plays in the context of another of Weinberg's creations, 
the model-independent framework of effective field theories.
\end{abstract}

\vspace{0.5cm}
\begin{center}
{\it Dedicated to the memory of Steven Weinberg, \\
who always chose the right degrees of freedom.}
\end{center}

\end{titlepage}

\section{Introduction}
\label{intro}

Steven Weinberg was one of those physicists who turned everything they touched
into gold --- except, of course, for their less well-known work.
Steve once or twice told me that few things gave him as much pleasure
as some recognition, such as a citation, of one of his more obscure
papers that he nevertheless was proud of. The context 
was his paper on the algebraic realization of 
chiral symmetry \cite{Weinberg:1969hw}, 
a beautiful and underappreciated idea 
related to some work we were doing in the early 1990s 
\cite{Weinberg:1990xm,Dicus:1992rw,Beane:1994ds}. 

Steve was not a conceited man, and he realized some of his work was not golden
--- even though by other people's standards it could very well seem so.
He has also told me --- and is even on record in some YouTube video I can no 
longer find --- that some of his papers in the 1960s were better
forgotten. I do not know that this is for sure the case, but he might 
have been referring to his paper on the evidence
that the deuteron is not an ``elementary'' particle
\cite{Weinberg:1965zz}, which followed a 
series of papers on ``quasiparticles'' 
\cite{Weinberg:1962hj,Weinberg:1963zza,Weinberg:1964zza}.
By 2007, this paper 
had been cited in only about 20 publications, according to inSpire. 
The only citation by Steve himself I could locate is
Ref. \cite{Weinberg:1995mt}, where the ``compositeness'' condition
$Z=0$ is not considered particularly ``useful'' in quantum field theories,
except  
that ``it fixes the coupling of the deuteron to the neutron and proton''.
When we were looking into the deuteron also in the 1990s 
\cite{VanKolck:1993ee,Ordonez:1993tn,Ordonez:1995rz},
not once did Steve recommend Ref. \cite{Weinberg:1965zz} to me. 
Yet, it has now 300+ citations (again, in inSpire)
due to the explosion of interest in the new exotic hadronic states
found since the $X(3872)$ in 2003 \cite{Belle:2003nnu}.
(For a recent review, see Ref. \cite{Brambilla:2019esw}.)
In this context, Steve's result has been invoked
as a criterion to decide which of the new states 
should be viewed as hadronic molecules. 

I believe 
Steve's loss of interest in that paper \cite{Weinberg:1965zz}
is due to his later, much
grander, idea of effective field theory (EFT) \cite{Weinberg:1978kz}, where
from the get-go experimental limitations
on accessible energies are explicitly incorporated in theory construction.
The notion of compositeness is tied to resolution:
one effective-field theorist's ``elementary'' particle is most
likely another's ``composite'' particle. 
In the nuclear EFTs formulated since the 1990s \cite{Hammer:2019poc}, 
Steve's criterion of compositeness is automatically satisfied.
The naive interpretation that an ``elementary'' particle 
is one associated with an explicit degree of freedom is muddled by the fact
one can always introduce an auxiliary field for the deuteron and/or 
another for the virtual spin-singlet $S$-wave state. 
If anything, it is for the latter \cite{Long:2013cya,SanchezSanchez:2017tws}
that the auxiliary field \cite{Kaplan:1996nv}
has been more useful
despite its intuitively higher ``compositeness'' coming from its extreme 
proximity to the scattering threshold. 
More generally, because an EFT includes {\it all} interactions
allowed by symmetries, the choice of fields is to some extent arbitrary
as far as observables, such as the location of $S$-matrix poles, go.

Nevertheless, identifying the relevant degrees of freedom 
is one of the cornerstones of EFT.
I have long had the feeling that Steve's 
criterion should have an EFT interpretation. I bumped into the paper
\cite{Weinberg:1965zz}
on my own as a graduate student and have reread it every 5-10 years since,
hoping to understand its EFT ramifications. 
Only recently did I succeed, thanks to 
a formulation of EFT for short-range forces \cite{Habashi:2020qgw} 
that is different
from the one \cite{vanKolck:1997ut,Kaplan:1998tg,Kaplan:1998we,vanKolck:1998bw}
that best applies to the two-nucleon problem.
Since EFT is the current paradigm for 
theory construction \cite{Weinberg:1996kw},
it seems worthwhile to relate Weinberg's compositeness to the choice
of degrees of freedom and organization of interactions (``power counting'').

This paper is not intended as a comprehensive review 
of the literature on Weinberg's compositeness, 
but instead should be viewed as a reconstruction of some 
known results from the point of view of a single-minded effective-field 
theorist. In Sec. \ref{shortEFT}, the short-range EFT relevant for a discussion
of Weinberg's compositeness criterion is presented and contrasted
with the short-range EFT normally considered in
the two-nucleon system.
The relevant EFT contains two $S$-wave poles, in terms of which Weinberg's 
relations are rewritten in Sec. \ref{poles}.
In Sec. \ref{dibaryon} the EFT is recast in terms of a ``diparticle''
field and Weinberg's $Z$ is shown to track the relative importance
of the diparticle kinetic term.
In both of these sections, the ``elementary'' limit is seen to be special,
with some of the results from Ref. \cite{Habashi:2020ofb} 
for narrow resonances extended to a bound/virtual state pair. 
The generalization of Weinberg's criterion to resonances is
discussed in Sec. \ref{narrow} based on the 
similarities between symmetric bound/virtual state and narrow resonances.
Section \ref{conc} concludes with some additional thoughts.

\section{EFT of short-range forces}
\label{shortEFT}

I will limit myself to the situation considered by 
Weinberg \cite{Weinberg:1965zz}, namely
a two-body process at characteristic 
momentum $Q$ much smaller than the mass $m$ of the
particles, $Q\ll m$, and the inverse of the force range $R$,
$Q\ll R^{-1}$. In this situation particles are nonrelativistic
and share a trivial factor $\exp(-imt)$ in their temporal evolution.
We introduce a field $\psi$ that involves only particle annihilation,
from which the trivial evolution factor is removed.
That is, we measure energies in the two-body system with respect to $2m$. 
Lorentz symmetry is implemented through ``reparametrization invariance''
\cite{Luke:1992cs}
and gives rise to an expansion of observables in powers of $Q/m$.
Particles cannot resolve the details of the underlying
interaction, which can be treated via a multipole
expansion as in electrodynamics:
operators in the Lagrangian involve multiple powers of $\psi$ and
$\psi^\dagger$ and their derivatives at the same point, and
generate an expansion of observables in powers of $QR$.
The coefficients of these operators --- ``Wilson'' or 
``low-energy coefficients'' (LECs) --- 
account for the details of the underlying
interaction and the effects of particle-antiparticle loops.
Moreover, I assume for simplicity conservation of particle
number ({\it i.e.}, the particle is stable), which means all operators
involve the same number of $\psi$s and $\psi^\dagger$s.
The theory separates into sectors of given particle number and 
here we can limit ourselves to operators with at most two $\psi$s.
In addition, to focus on the essential
points, I impose parity and time-reversal invariance,
and neglect spin and internal quantum numbers such as isospin.

Under these conditions the most general Lagrangian can be written as
\cite{vanKolck:1998bw,Habashi:2020qgw}
\begin{eqnarray}
\mathcal{L} &=& 
\psi^{\dagger} \left(i\frac{\partial}{\partial t}
+\frac{\overrightarrow{\nabla}^2}{2 m}+ \ldots \right)\psi -\frac{4\pi}{m}
\biggl\{C_{0} \left( \psi \psi \right)^{\dagger} \left( \psi \psi \right) 
- \frac{C_{2}}{8}\left[\left( \psi \psi \right)^{\dagger} 
\left(\psi \overleftrightarrow{\nabla}^2 \psi \right) + \mathrm{H.c.} \right]
\nonumber\\
&&
+\frac{C_{4}}{64}
\left[\left( \psi \psi \right)^{\dagger}
\left(\psi \overleftrightarrow{\nabla}^4\psi \right) + \mathrm{H.c.}
+ 2\left( \psi\overleftrightarrow{\nabla}^2\psi \right)^{\dagger}
\left( \psi\overleftrightarrow{\nabla}^2\psi \right) \right]
+ \ldots 
\biggr\}, 
\label{Lag}
\end{eqnarray}
where 
$\overleftrightarrow{\nabla} \equiv 
\overrightarrow{\nabla}-\overleftarrow{\nabla}$
and $C_{2n}$ are real LECs.
The ``$\ldots$'' represent 
operators that only contribute at higher orders and/or higher waves.

Using the rules of quantum field theory we can calculate the $T$ matrix for 
two-body scattering. 
At sufficiently low energy $E\equiv k^2/m$ in the 
center-of-mass (CM) frame, we expect $S$ waves to dominate.
The corresponding $S$ matrix can be written
\begin{equation}
S_{0}(k)=\exp(2i\delta_0(k))=1-\frac{imk}{2\pi}T_{0}(k)
\label{S0}
\end{equation}
in terms of the $S$-wave phase shift $\delta_0(k)$ and on-shell
$T$ matrix $T_{0}(k)$.
Because we are considering only
short-range interactions, the EFT $T$ matrix
is related to the effective range expansion (ERE) \cite{Bethe:1949yr},
\begin{equation}
T_0(k)=-\frac{4\pi}{m}\left(k\cot\delta_0(k)-ik\right)^{-1} 
=-\frac{4\pi}{m}\left[-\frac{1}{a_{0}} + \frac{r_{0}}{2}\,k^2 
  - P_{0} \left(\frac{r_{0}}{2}\right)^3 k^4 + \ldots - i k \right]^{-1},
\label{T0}
\end{equation}
where $a_0$, $r_0$, $P_0$, {\it etc.} are, respectively,
the scattering length, effective range, shape parameter, {\it etc.}
(Note that I use the original convention for the sign of the scattering length.)
While at the two-body level short-range EFT can only amount to a 
reorganization of the ERE, 
it does allow for a consistent description of three and more particles
\cite{Hammer:2019poc}.

What reorganization of the ERE is needed depends on the magnitudes
of the ERE parameters in Eq. \eqref{T0}
or, equivalently, of the LECs in Eq. \eqref{Lag} connected to them 
by renormalization.
Contact interactions are singular and require regularization,
as in any other EFT.
Since we are in general interested in a nonperturbative problem,
a momentum regulator is more useful than dimensional regularization.
I will denote the regulator cutoff parameter by $\Lambda$
and the regulator function by $F(q^2/\Lambda^2)$,
with $F(0)=1$ and $F(x^2\gg 1)\ll 1$. 
Regulator dependence enters through odd powers, $\Lambda^{2n+1}$,
each accompanied by a regulator-dependent number
\begin{equation}
\theta_{2n+1}\equiv \frac{2}{\pi}\int_0^\infty\!\!dx\,x^{2n}F(x^2).
\label{thetas}
\end{equation}
For example, 
$\theta_{2n+1}=2/(2n+1)\pi$ for a sharp-cutoff regulator, 
$F(q^2/\Lambda^2)=\theta(1-q/\Lambda)$.
Renormalization ensures that at any order regulator artifacts
appear in observables only through terms with positive powers of $Q/\Lambda$.
After renormalization,
residual regulator effects are no larger
than the relative error ${\cal O}(QR)$ arising from the truncation of the
Lagrangian once the cutoff parameter is sufficiently large,
$\Lambda\simge R^{-1}$.

Order-by-order renormalization requires that the bare parameters
in the Lagrangian \eqref{Lag} be specific functions of the cutoff
which cancel the cutoff dependence coming from the loops.
The critical issue for an EFT is the magnitude of the surviving parts,
which do contribute to observables. They can be estimated with the principle
of naturalness \cite{tHooft:1979rat,Veltman:1980mj}: the
sensitivity to high-momentum physics reflected in the cutoff
dependence of the bare LECs is captured in the surviving parts by
the replacement $\Lambda \to R^{-1}$. 

In particle physics, naturalness is usually employed in a perturbative
context. Here, adding to an existing diagram 
a (nonrelativistic) loop with an extra $C_{2n}$ interaction
brings in a factor of $l^{3}/4\pi$ from the integration measure,
$m/l^2$ from the two-particle propagator, and $4\pi C_{2n}l^{2n}/m$ 
from the vertex, where $l$ is the loop momentum.
These factors combine for a net $C_{2n}l^{2n+1}$, which increases
the cutoff dependence by $C_{2n}\Lambda^{2n+1}$.
Naturalness then suggests $C_{2n}={\cal O}(R^{2n+1})$. 
This, in turn, implies that after renormalization the
series is a perturbative expansion in $QR$ \cite{vanKolck:1998bw},
justifying looking at the addition of a single loop in the first place. 
In first nontrivial order we have the tree graph with $C_0$, at next
order the one-loop graph with two $C_0$ vertices, and so on,
corresponding to an expansion of Eq. \eqref{T0} where 
$a_0^{-1}$ is the largest term,
$ik$ the second largest, $r_0k^2/2$ the next largest, and so on. 
Higher waves start at third nontrivial order.
This notion of naturalness incorporates
the simpler idea that, in the absence of a specific arrangement
of parameters in an underlying potential, 
all ERE parameters are given by $R$ according
to their dimensions, for example $|a_0|\sim |r_0|/2\sim R$ and $|P_0|\sim 1$. 
An example is 
provided by the spherical well \cite{vanKolck:1998bw}.
Note the difference to naturalness applied
to relativistic loops, where we normally
estimate the measure as $l^{4}/(4\pi)^2$ and the two-particle
propagator as $l^{-2}$, leading to the so-called naive dimensional analysis
\cite{Manohar:1983md,Georgi:1992dw}. 

Unfortunately, in the natural case there are no $S$-matrix poles within
the range of validity of the EFT. To produce shallow poles, at least
one $C_{2n}$ must have an unnaturally large part and demand a
nonperturbative treatment at LO.  In the two-nucleon
system each $S$ wave, $^3S_1$ and $^1S_0$, contains a single shallow pole,
respectively a bound state (the deuteron) and a virtual state. In
an $S$-wave channel with a single shallow pole, 
there is a low-energy scale $\aleph \ll R^{-1}$ corresponding
to the magnitude of the binding momentum. This situation
can be described in the EFT if $C_0={\cal O}(\aleph^{-1})$ and
thus $|a_0|\sim \aleph^{-1}$,
while the remaining $S$-wave LECs scale as $C_{2n>0}={\cal O}(\aleph^{-2}R^{2n-1})$ 
\cite{vanKolck:1998bw}. In this case, an additional insertion of $C_0$ incurs in
a penalty of $Q/\aleph$, which demands resummation with
a pole at $Q\sim \aleph$ within the EFT range of validity.
The leading order (LO) consists of the exact solution of the Lippmann-Schwinger
(or Schr\"odinger) equation for the 
momentum-space potential
\begin{equation}
V_0^{(0)}(\Lambda)= \frac{4\pi}{m}\, C_0^{(0)}(\Lambda),
\label{pot1pole}
\end{equation}
which can be renormalized as long as the LO piece of $C_0$ is
\begin{equation}
C_{0}^{(0)}(\Lambda) =
-\frac{1}{\theta_{1}\Lambda}
\left[1 + \delta + \mathcal{O}\left(\delta^2\right) \right], 
\label{C0}
\end{equation}
where 
\begin{equation}
\delta \equiv \frac{1}{\theta_{1}a_{0}\Lambda}
\label{delta}
\end{equation}
because I chose to reproduce $a_0$ exactly at LO.
Meanwhile, corrections from $C_{2n>0}$ remain perturbative in $QR$,
with higher ERE parameters still taking natural values, 
for example $|r_0|\sim R$ and $|P_0|\sim 1$. 
At next-to-leading order (NLO) we have one insertion of $C_2$,
at next-to-next-to-leading order (N$^2$LO) two insertions of $C_2$,
and so on, with higher waves unchanged and thus starting
with $P$ at N$^3$LO.
This unnatural case requires a fine tuning in the 
underlying potential, for example a precise adjustment of the depth 
of a spherical well to
produce a shallow $S$-wave pole \cite{vanKolck:1998bw}.
For a discussion of naturalness in a nonperturbative context, see Ref. 
\cite{vanKolck:2020plz}.

The case relevant for a discussion of Weinberg's compositeness requires
both $C_0={\cal O}(\aleph^{-1})$ and $C_2={\cal O}(\aleph^{-3})$ to be
enhanced. Both interactions need to be iterated at LO, which is equivalent to
an exact solution of
\begin{equation}
V_0^{(0)}(\vec{p}\,',\vec{p};\Lambda)= \frac{4\pi}{m} \left[C_0^{(0)}(\Lambda) 
+ C_2^{(0)}(\Lambda)\frac{\vec{p}\,^{2}+\vec{p}\,'^{2}}{2}\right],
\label{pot2poles}
\end{equation}
written in terms of the incoming and outgoing relative momenta $\vec{p}$ 
and $\vec{p}\,'$, respectively.
It is not obvious that the increased singularity of the two-derivative
contact interaction allows for renormalization, and yet it does if the LO 
bare LECs are \cite{Phillips:1997xu,Beane:1997pk,Habashi:2020qgw}
\begin{eqnarray}
C_{0}^{(0)}(\Lambda) & = &
\frac{\theta_{5}}{\theta_{3}^{2} \Lambda}
\left[1 \mp 2\varepsilon
+\left(1-\frac{\theta_{3}^{2}}{\theta_{5}\theta_{1}}\right)\varepsilon^2 
\pm \left(1- \frac{\theta_{3}\theta_{-1}}{\theta_{1}^{2}} 
- \frac{\theta_{3} r_0}{\theta_{1}^{3} a_{0}}\right)\varepsilon^3
+ \mathcal{O}\left(\varepsilon^4, \frac{r_0}{a_0}\varepsilon^4\right) 
\right],
\label{bareC0} \\
C_{2}^{(0)}(\Lambda)  & = &-\frac{2}{\theta_{3}\Lambda^3} 
\left[1 
\mp \varepsilon
\pm \left(1- \frac{\theta_{3}\theta_{-1}}{\theta_{1}^{2}} 
- \frac{\theta_{3} r_0}{\theta_{1}^{3} a_{0}}\right)\frac{\varepsilon^3}{2}
+ \mathcal{O}\left(\varepsilon^5, \frac{r_0}{a_0}\varepsilon^5\right) 
\right],
\label{bareC2}
\end{eqnarray}
where
\begin{equation}
\varepsilon \equiv 
\left(-\frac{2 \theta_{1}^{2}}{\theta_{3}r_{0}\Lambda}\right)^{1/2}.
\label{varepsilon}
\end{equation}
While $C_0^0(\Lambda)$ is $\propto \Lambda^{-1}$ like in Eq. \eqref{C0}, now
there are two solutions indicated by the $\pm$ signs
accompanying the unusual inverse powers of $(- r_{0} \Lambda)^{1/2}$.
This square-root means that we must have $r_{0} < 0$ for real $C_{0,2}^{(0)}$.
This result is consistent with the Wigner bound
\cite{Wigner:1955zz}, which puts a condition on the rate of 
change of the phase shift with respect to the energy for a finite-range,
energy-independent potential and implies \cite{Fewster:1994sd,Phillips:1996ae}
\begin{equation}
r_{0} \leq 2R \left(1 - \frac{R}{a_{0}} + \frac{R^2}{3a_{0}^2}\right)\, .
\label{r0}
\end{equation}
For a zero-range potential, 
there is no positive effective range consistent with Eq.~\eqref{r0}
\cite{Beane:1997pk}. 
The requirement of renormalizability for our LO interactions
automatically yields the same constraint and leads
to the situation considered by Weinberg. 

With this power counting, the EFT 
results not only in $|a_0|\sim \aleph^{-1}$ but also 
$|r_0|\sim \aleph^{-1}$.
There are now two poles within the EFT domain.
They can be: 
\begin{itemize}
\item for $a_0<0$,
either two resonance poles in the lower half of the complex $k$ plane
which are symmetric with respect to the imaginary axis (when $|a_0|<2|r_0|$),
or two (separate or coincident) poles on the negative imaginary axis
(when $|a_0|\ge 2|r_0|$).
\item for $a_0>0$,
one pole on the negative and the other on the positive imaginary axis.
\end{itemize}

At NLO, $C_4={\cal O}(\aleph^{-4}R)$ enters, causing a small
shift in pole positions if $a_0$ and $r_0$ are kept fixed,
and generating a shape parameter $|P_0|\sim \aleph R$.
Naturalness applied to the error at NLO implies \cite{Habashi:2020qgw} 
that $C_6={\cal O}(\aleph^{-5}R^2)$ appears at N$^2$LO, and it
was speculated that $C_{2n>0}={\cal O}(\aleph^{-2n}R^{n-1})$.
This situation requires a potential with two fine-tuned parameters, an
example being a repulsive delta shell added at the rim
of a spherical well \cite{Gelman:2009be,Habashi:2020qgw}.

The three cases --- natural $S$ waves, one fine tuning, and 
two fine tunings --- are contrasted in Table \ref{3efts}.
As elaborated below, the case of interest for Weinberg's argument
involves two poles.
The emphasis in Ref. \cite{Habashi:2020qgw} was on resonances --- in fact, 
resonances in the sense of Weinberg \cite{Weinberg:1963zza}, 
meaning poles of the $S$ matrix regardless of
the existence of a sharp peak in the cross section.
Resonances were revisited with an auxiliary resonance field in
Ref. \cite{Habashi:2020ofb}, elaborating on earlier work
\cite{Bedaque:2003wa,Gelman:2009be,Alhakami:2017ntb}. 
I will consider in more detail here the case where one of the two poles
is a bound state.

\begin{table}[tb]
\begin{center}
\begin{tabular}{|c||c|c|c|}
\hline
\# $S$-wave poles  & 0 
        & 1
        & 2
\\
\hline
\hline
LO      & 0
        & $C_0$
        & $C_{0,2}$
\\
NLO     & $C_0$
        & $C_2$
        & $C_4$ 
\\ 
N$^2$LO & $C_0^2$ 
        & $C_2^2$ 
        & $C_4^2$, $C_6$
\\
\hline
\end{tabular}
\end{center}
\caption{Schematic representation of interaction
orderings in short-range EFT  with different number of shallow $S$-wave poles. 
At each of the three lowest orders
--- leading (LO), next-to-leading (NLO), and 
next-to-next-to-leading (N$^2$LO) ---
the relevant low-energy constants (LECs) $C_{2n}$ are indicated.
For an LO LEC, all powers are understood.
\label{3efts}}
\end{table}

\section{Poles}
\label{poles}

The distinctive feature of the short-range EFT 
with two fine tunings is the existence of two poles $k_{\pm}$
for the $S$-wave $S$-matrix,  
\begin{equation}
S_{0}(k)= e^{2i\phi(k)} \frac{(k+k_-)(k+k_+)}{(k-k_-)(k-k_+)},
\label{eq.39}
\end{equation}
where $\phi(k)$ is the nonresonant or ``background'' contribution 
to the phase shift.
For $a_{0}>0$ the two poles are on opposite sides of the imaginary
axis \cite{Habashi:2020qgw}, and at LO
\begin{equation}
k_{\pm}^{(0)} = \mp i \kappa_{\pm} , 
\label{polesLO}
\end{equation}
where
\begin{equation}
\kappa_\pm = \frac{1}{|r_{0}|} 
\left(\sqrt{1+\frac{2|r_{0}|}{a_{0}}} \pm 1\right),
\qquad
\kappa_+>\kappa_->0.
\label{kappasLO}
\end{equation}

The residues of $iS_0^{(0)}$ at both poles obey 
\cite{Hyodo:2013iga,Habashi:2020qgw}
\begin{equation}
1-Z \equiv 
\frac{1}{2\kappa_\pm}
\left.\mathrm{Res}\!\left(iS_0^{(0)}\right)\right|_{\mp i \kappa_\pm} 
=\frac{\kappa_{+} - \kappa_{-}}{\kappa_{+} + \kappa_{-}}
=\left(1+2\frac{|r_{0}|}{a_{0}}\right)^{-1/2},
\qquad 
1> Z >0.
\label{Z}
\end{equation}
The positive value indicates
that the pole $k_-$ on the positive imaginary axis 
is a bound state \cite{Moller:1946}.
The constraint $r_0<0$ from renormalization thus excludes \cite{Habashi:2020qgw}
the possibility of a ``redundant'' pole 
\cite{Ma:1946,TerHaar:1946,Ma:1947zz}
on the positive imaginary axis with negative residue. 

Weinberg's $Z$ parameter, as defined for this type of EFT,
is thus an observable, essentially
the inverse relative splitting of the two states \cite{Baru:2003qq}.
It allows us to rewrite other quantities in terms of $Z$ 
and the binding momentum $\kappa_-$:
\begin{equation}
\kappa_+ = \kappa_- \frac{2-Z}{Z}
\label{kappa+LO}
\end{equation}
and
\begin{eqnarray}
a_0 &=& \frac{1-Z}{1-Z/2} \, \kappa_-^{-1},
\\
r_0 &=& 
-\frac{Z}{1-Z} \, \kappa_-^{-1},
\end{eqnarray}
which are the Weinberg relations, valid up to corrections 
of ${\cal O}(\aleph R)$
\cite{Weinberg:1965zz}. 
At LO, $\phi^{(0)}(k)=0$ and
\begin{equation}
k\cot \delta^{(0)}(k)= 
-\kappa_-\left[1+\frac{Z}{2(1-Z)}\left(1+\frac{k^2}{\kappa_-^2}\right)\right].
\label{cotdeltaLO}
\end{equation}
The appearance of a new parameter at NLO obviously implies modifications
of these relations, as there are then three independent parameters
instead of two. For the effects of an additional parameter,
see Ref. \cite{Albaladejo:2022sux}.

When the bound state is within the regime of the EFT, 
$\kappa_-\sim \aleph \ll R^{-1}$, both scattering length and
effective range are large compared to $R$ unless 
$Z\to 0$ or $Z\to 1$. 
These limits deserve special consideration,
as additional small parameters lead to a reorganization of the
hierarchy of interactions.
When $Z$ is within the expansion parameter $\kappa_- R$ of $0$ or $1$,
some contributions we have considered LO so far are demoted
to higher orders.

The case $Z\simle \kappa_- R$
corresponds to a natural
effective range, $|r_0|\simle R$, and unnaturally large scattering length,
$a_0\sim \kappa_-^{-1}$. We recover the situation
dealt with for general short-range potentials in  
Ref. \cite{vanKolck:1998bw} where the virtual state is outside 
the EFT regime. Range effects, contained in the
second term in Eq. \eqref{cotdeltaLO}, are included in perturbation theory.
How well the EFT works for the deuteron can be seen in 
Ref. \cite{Chen:1999tn}.
It has had much use in nuclear \cite{Hammer:2019poc}
and cold atomic systems \cite{Braaten:2007nq}.
This is the case Weinberg \cite{Weinberg:1965zz}
deemed ``composite'', with $1-Z$ being then the ``compositeness'' 
of the bound state. 

Since multiple shallow poles are associated 
with $|r_0|\gg R$, Morgan \cite{Morgan:1992ge} 
suggested that their existence is 
likely an indication of ``elementarity''.
As Ref. \cite{Morgan:1992ge} admits, however, there {\it are} potentials
with two poles. A local class of such potentials, with a perhaps unphysical
exponential fall-off at large distances, was given long ago
\cite{Bargmann:1949}. For more abrupt fall-off, it does seem
that an increasing number of shallow poles requires additional fine tuning. 
In contrast, it might be reasonable to expect that 
the strong dynamics of two particles coupled
to an elementary ``diparticle'' with the two-particle quantum numbers 
could give rise to multiple poles.
Even in this case, however,
the close proximity of the diparticle mass 
to $2m$ would still presumably require explanation --- if there is
no symmetry, then likely we would be back to fine tuning.
Without further information, the mere existence of two shallow
poles is far from certain fulfillment of
an intuitive notion of ``elementariness''. 

However, the case 
$1-Z\simle \kappa_- R$ does seem special.  
Up to ${\cal O}(\kappa_- R)$ corrections, it corresponds
to two poles $\kappa_+= \kappa_-$
close to the origin and a large effective range \cite{Weinberg:1965zz} 
$|r_0| \simge 1/(\kappa_-^2 R)$, 
while the scattering length is limited,
$a_0 \simle 2R$. For real $k^2$ not too close to the origin, scattering
is dominated by the second term in Eq. \eqref{cotdeltaLO}.
In fact, for $k\sim \kappa_-$, it is ${\cal O}((\kappa_- R)^{-1})$
and larger than the unitarity term $-ik$ in $T_0(k)$. 
Like for narrow resonances \cite{Habashi:2020qgw},
this poses a problem for power counting:
a subleading unitarity term means perturbation theory, but at tree level
the sum of the two interactions with LECs $C_0$ and $C_2$ supports
no pole. 

The way to get two poles very close to the real axis is to assume that
there is a correlation between an infinite number of interactions
in Eq. \eqref{Lag} such that at leading nontrivial
order they add to a nonlocal potential
such as
\begin{equation}
V_0^{(1)}(\vec{p}\,',\vec{p})= \frac{4\pi}{m} 
C_0^{(1)}\frac{1}{\sqrt{1-C_2^{(1)}\vec{p}\,'^{2}/C_0^{(1)}}}
\frac{1}{\sqrt{1-C_2^{(1)}\vec{p}\,^{2}/C_0^{(1)}}},
\label{pot2sympoles}
\end{equation}
instead of Eq. \eqref{pot2poles}.
For very different reasons,
this type of potential has recently been considered as an alternative
LO for the two-nucleon system at physical \cite{Beane:2021dab}
(also with explicit pions \cite{Peng:2021pvo})
and unphysical \cite{Timoteo:2022}
quark masses. Here, it is easy to see that it yields Eq. \eqref{cotdeltaLO} if 
\begin{eqnarray}
C_{0}^{(1)} & = & a_0,
\label{bareC0sympoles} \\
C_{2}^{(1)}  & = & a_0 \, \kappa_-^{-2}.
\label{bareC2sympoles}
\end{eqnarray}
Since $C_0^{(1)}= {\cal O}(R)$ and 
$C_2^{(1)}= {\cal O}(R \aleph^{-2})$,
the potential is ${\cal O}(R)$
and perturbative for generic momenta ${\cal O}(\aleph)$, 
but it produces the correct poles. 
They are poles of the potential itself, not of the
iteration of the potential to all orders as one would expect 
from a ``composite'' state. This is the simplest rationale 
for the interpretation of $Z\to 1$ as the ``elementary''-particle limit.

The price we pay in the EFT without a field for the diparticle
is a loss of locality. Locality is usually taken
as one of the principles in EFT construction. By locality I mean
not that interactions are momentum independent --- they cannot be
forbidden by symmetries --- but that at any order only a finite 
number of these interactions is present. As we will see next,
this intrinsic nonlocality can be traded for energy dependence.

\section{Diparticle}
\label{dibaryon}

In an EFT, the choice of fields is arbitrary. We may, in particular,
introduce an auxiliary ``diparticle'' (``dibaryon'' or ``dimer'' in
nuclear or atomic physics) field $d$ \cite{Weinberg:1962hj,Kaplan:1996nv}
with the quantum numbers of the bound state.
Consistently with our choice for $\psi$, 
$d$ is a heavy field from which the trivial evolution $\exp(-2imt)$ 
has been removed, leaving behind a residual mass $\Delta$.
The most general Lagrangian can be written as 
\cite{Kaplan:1996nv,Habashi:2020ofb}
\begin{eqnarray}
\Lag & = & \psi^\dagger \left(i\frac{\partial}{\partial t} 
+ \frac{\vec{\nabla}^2}{2 m} \right) \psi 
+ d^\dagger \left( i\frac{\partial}{\partial t} 
+ \frac{\vec{\nabla}^2}{4 m} - \Delta \right) d 
\notag \\ 
& & 
+ \sqrt{\frac{4 \pi}{m}} \, \frac{g_{0}}{4} 
\left( d^\dagger \psi \psi + \psi^\dagger\psi^\dagger d \right)  
- \frac{4 \pi}{m} \tilde{C}_{0} 
\left(\psi \psi\right)^\dagger\left(\psi \psi\right)
+ \ldots ,
\label{Ldimer}
\end{eqnarray}
where 
$\tilde{C}_{0}$ and $g_{0}$ are real LECs
and ``$\ldots$'' indicates terms with additional derivatives
(and fields, were we considering more-body systems). 
There is a certain redundancy in the Lagrangian \eqref{Ldimer},
since when one integrates out $d$ one obtains a string of
four-$\psi$ interactions of the form already present in $\Lag$
\cite{Bedaque:1999vb}.
However, these terms are all correlated in a way that is explicit
only once the dimer field is kept. 
LECs such as $\tilde{C}_{0}$ can be thought of as the uncorrelated part of 
these interactions.

\begin{figure}[t]
\centering
\includegraphics[trim=0.00cm 0.00cm 10.00cm 26.00cm,clip]{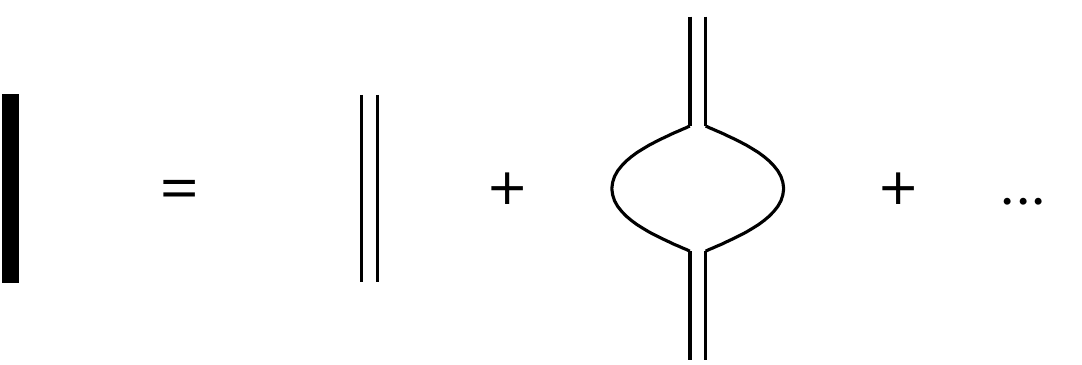}
\caption{Full diparticle propagator at
LO. Solid lines represent particles and a double line, a bare diparticle.
\label{fulldpropfig}}
\end{figure}

Now two particles can interact, through the coupling $g_0$,
by the exchange of the diparticle in the $s$ channel. 
The bare diparticle propagator in its CM frame is
\begin{equation}
D^{(0)}_0(k,\Lambda) = \left[k^2/m - \Delta^{(0)}(\Lambda) + i\epsilon\right]^{-1}
\label{baredprop}
\end{equation}
in terms of the LO residual mass $\Delta^{(0)}$.
This propagator has two poles on the imaginary $k$ axis for $\Delta^{(0)}<0$,
but they are equidistant from the origin. 
For them to be shallow, $\Delta^{(0)} = {\cal O}(\aleph^2/m)$.
The splitting between $\kappa_+$ and $\kappa_-$, and thus $1-Z$,
comes from the dressing of the propagator
with loops. The first quantum effect 
in the bare diparticle propagator \eqref{baredprop}
is the one-loop diparticle self-energy 
\begin{equation}
\Pi^{(0)}(k,\Lambda)=  g_0^{(0)2} \, I_0(k,\Lambda) ,
\label{loopdprop}
\end{equation}
where $g_0^{(0)}$ is the LO diparticle-particle coupling
and 
\begin{equation}
I_0(k,\Lambda) = 4\pi\int\!\!\frac{d^3l}{(2 \pi)^3}\,
\frac{1}{l^2 - k^2 -i \epsilon} 
= \theta_{1} \Lambda +  i k + \theta_{-1} k^2/\Lambda +\ldots 
\label{I0}
\end{equation}
is the two-particle bubble.
If $g_{0}^{(0)}={\cal O}(\sqrt{\aleph/m})$ \cite{Habashi:2020ofb}, 
at LO we need to resum the diparticle self-energy 
as in Fig. \ref{fulldpropfig}, resulting in the dressed dimer propagator
\begin{equation}
D^{(0)}(k) = \left[D^{(0)-1}_0(k,\Lambda) + \Pi^{(0)}(k,\Lambda)\right]^{-1}
\label{dresseddprop}
\end{equation}
and the LO amplitude
\begin{equation}
T_0^{(0)}(k) =  \frac{4 \pi}{m}\,g_{0}^{(0)2}D^{(0)}(k) 
=-\frac{4 \pi g_{0}^{(0)2}}{m}
\left[\Delta^{(0)}(\Lambda)- k^2/m- \Pi^{(0)}(k,\Lambda)\right]^{-1}.
\label{T0dimer}
\end{equation}

The poles $\pm i\kappa_\mp$ of $T_0^{(0)}(k)$ obviously satisfy
\begin{equation}
\Delta^{(0)}(\Lambda)+\kappa_\mp^2/m- \Pi^{(0)}(\pm i\kappa_\mp,\Lambda)=0.
\label{polesdimer}
\end{equation}
Since $B_{-}=\kappa_-^2/m$ is the binding energy of the bound state,
$\Delta^{(0)}(\Lambda)$ can be viewed as the bare or ``unrenormalized'' 
energy of the state.
As pointed out by Weinberg \cite{Weinberg:1962hj}, this relation does
not allow us to calculate $B_{-}$; it only relates bare and renormalized
energies.
The essential cutoff dependence of the self-energy can be removed
through the renormalization conditions
\begin{eqnarray}
\Delta^{(0)}(\Lambda) & = & -\frac{2}{m r_{0}}
\left(\theta_{1} \Lambda-\frac{1}{a_{0}}\right)
=\frac{2\kappa_-}{m}\,\frac{1}{Z}
\left[(1-Z)\theta_{1} \Lambda-(1-Z/2) \kappa_-\right],
\label{Delta0Lambda} \\
g_{0}^{(0)} & = & \left(-\frac{2}{m r_{0}}\right)^{1/2}
=\left(\frac{1-Z}{Z}\,\frac{2\kappa_-}{m}\right)^{1/2}, 
\label{g0Lambda}
\end{eqnarray}
which lead again to Eqs. \eqref{cotdeltaLO}
and \eqref{kappasLO}.
The scalings $\Delta^{(0)}= {\cal O}(\aleph^2/m)$ and 
$g_{0}^{(0)}= {\cal O}(\sqrt{\aleph/m})$ when $Z$ is not particularly close
to 0 or 1 are evident from the finite pieces, respectively $\propto B_-$ and 
$\propto (B_-)^{1/4}$ \cite{Weinberg:1962hj}.
The negative finite part of $\Delta^{(0)}(\Lambda)$ ensures the two poles 
have imaginary momentum. 

The residual cutoff dependence $\propto \Lambda^{-1}$ 
of the LO amplitude \eqref{T0dimer} implies that there are corrections
at NLO. In addition to shifts $\Delta^{(1)}(\Lambda)$ and $g_{0}^{(1)}(\Lambda)$ 
in parameters that appear already at LO,
renormalization requires the inclusion of $\tilde{C}_{0}^{(1)}={\cal O}(R)$,
which is determined by the shape parameter $P_0$, $|P_0|\sim \aleph R$
\cite{Habashi:2020ofb}.
This is the EFT sense in which a nonperturbative 
interaction ($C_{0}={\cal O}(\aleph^{-1})$ in the previous section) 
becomes weak enough ($\tilde{C}_{0}={\cal O}(R)$) 
for perturbation theory to apply, thanks to the introduction of a field
with the quantum numbers of a bound state 
\cite{Weinberg:1962hj,Weinberg:1963zza}.
This EFT has been applied to the deuteron in Ref. \cite{Beane:2000fi}
and some of its atomic ramifications have been discussed 
in Ref. \cite{Braaten:2007nq}.

The historical interpretation of the limit $Z\to 0$ as compositeness seems 
to be more closely related to this formulation of the EFT.
It is tempting to assign compositeness to this limit on the basis
of the $Z^{-1}$ behavior of $\Delta^{(0)}(\Lambda)$.
After all, one might expect
compositeness to correspond to a large diparticle energy \cite{Weinberg:1962hj}
and dispense with the need for a field to represent the bound state.
However, $\Delta(\Lambda)$ is not an observable and the use
of the dimer field is not tied to $Z\simle \kappa_- R$.
In contrast, $g_{0}^{(0)}$ does not run with 
$\Lambda$ and is also large for $Z\simle \kappa_- R$ because
$|r_0|$ is small: from Eq. \eqref{g0Lambda} we see that
\begin{equation}
g_{0}^{(0)2}  \le \frac{2\kappa_-}{m Z},
\label{g0Lambdaprime}
\end{equation}
thus diverging as $Z\to 0$.
This relation
supports the intuitive interpretation
of compositeness as resulting from a strong interaction 
of the two-particle state with the diparticle \cite{Weinberg:1962hj}.
But again, $g_{0}(\Lambda)$ is not an observable.

Bruns \cite{Bruns:2019xgo,Bruns:2022hmb}
has provided insight into the discussion of Weinberg's compositeness
by looking at wavefunctions for specific models. 
It is seen that $1-Z$ is approximately
the probability to find the two particles
beyond the range of the potential.
Illuminating as it is, this sort of approach shares 
the limitation of others, which
focus on non-observable quantities. Only the long-range properties
of the wavefunction are related to observables; the short-range
behavior, which affects the probability
to find particles widely separated via the normalization of
the wavefunction, is intrinsically tied to two-body currents.
This is reflected on the wavefunctions
obtained in EFT. It is not difficulty to see that the LO wavefunction
for the bound state is 
\begin{equation}
\psi^{(0)}(\vec{r})=\sqrt{\frac{\kappa_-}{2\pi}}\frac{e^{-\kappa_- r}}{r}
\label{wfeq}
\end{equation}
in coordinate space. Not surprisingly,
this is the asymptotic behavior of the wavefunctions considered by
Bruns \cite{Bruns:2019xgo,Bruns:2022hmb}. However, since
the wavefunction \eqref{wfeq} extends to the origin, 
the probability of finding the particles outside the potential is 1.
This was to be expected, since at LO in the EFT the interaction has zero range.

These historical observations do, however, reflect a change
in the power counting when $Z\to 0,1$.
For $Z\simle \kappa_- R$,
$\Delta^{(0)}={\cal O}(\aleph/mR)$ and 
$g_{0}^{(0)}={\cal O}(1/\sqrt{mR})$. 
The $k^2/m$ in Eq. \eqref{T0dimer} is subleading by
a power of $\aleph R$ compared to the other terms.
This means the diparticle's kinetic energy enters only at NLO
\cite{vanKolck:1998bw}
to account for energy dependence,
while the scattering length is produced by the 
diparticle's residual mass.
The consequence of Weinberg's condition for $Z$ not particularly close to 0
is thus not the need for a diparticle field {\it per se}, 
but for its kinetic term to enter at LO.

More dramatic changes take place for $1-Z\simle \kappa_- R$,
when the coupling gets reduced to 
$g_{0}^{(1)}= {\cal O}(\sqrt{\aleph^2R/m})$ and the whole diparticle self-energy
is an NLO effect. 
We are in a perturbative regime where 
$\Delta^{(0)}(\Lambda)=\Delta^{(0)}<0$ gives rise at LO to two nearby,
shallow poles $\kappa_+= \kappa_-\equiv\kappa =\sqrt{m|\Delta^{(0)}|}$ 
from the bare particle propagator \eqref{baredprop}.
Because in this limit the diparticle behaves like a weakly interacting
particle of mass $\simeq 2m + \Delta^{(0)}$ just below the two particle
threshold, it is indeed natural to interpret it as an ``elementary''
particle. This relates to the usual interpretation of $Z$ as the overlap
between ``dressed'' and ``bare'' diparticle states \cite{Weinberg:1962hj}.
Again, one might expect some fine tuning to produce
$|\Delta^{(0)}|\ll 2m$, but in this situation a single fine tuning would
be needed.

This interpretation is supported by the diparticleless formulation.
When multiplied by $4\pi g_0^{2}/m$, 
the diparticle propagator \eqref{baredprop} 
can be seen as an energy-dependent potential, 
which replaces the diparticleless nonlocal potential \eqref{pot2sympoles}. 
The $s$-channel exchange of the diparticle automatically accounts for 
the correlations needed in the diparticleless formulation,
and the EFT remains local. However, as pointed out
in Ref. \cite{Gelman:2009be} in the closely related case of narrow resonances,
$g_0^{(1)2}D_0^{(0)} ={\cal O}(R)$ should now be comparable to 
the ``uncorrelated'' part of the short-range interaction,
$\tilde{C}_0^{(1)}={\cal O}(R)$ on the basis of naturalness. 
In the $Z\to 1$ limit we thus
expect the potential to be not just the diparticle propagator but
\begin{equation}
V_0^{(1)}(k)= \frac{4\pi}{m}
\left[\tilde{C}_0^{(1)} + \frac{mg_0^{(1)2}}{k^2 - m\Delta^{(0)}}\right].
\label{pot2sympolesendep}
\end{equation}
This potential has been discussed for the two-nucleon $^1S_0$ channel
in Ref. \cite{SanchezSanchez:2017tws}
and for narrow resonances in Ref. \cite{Habashi:2020ofb},
elaborating on the earlier work of Refs. 
\cite{Bedaque:2003wa,Gelman:2009be,Alhakami:2017ntb}. 
An analogous argument justifies the addition of a constant to the nonlocal
potential \eqref{pot2sympoles} in the diparticleless formulation.

Many of the resonance results can immediately be translated to the case of a 
symmetric bound/virtual state pair. In particular, the LO amplitude
\eqref{T0dimer} becomes
\begin{equation}
T_0^{(1)}(k) =  V_0^{(1)}(k)
=\frac{4 \pi}{m} a_0 \frac{1-k^2/k_0^2}{1+k^2/\kappa^2}, 
\label{T0elementarydimer}
\end{equation}
with
\begin{eqnarray}
\Delta^{(0)}& = & -\frac{\kappa^2}{m},
\label{elementaryDelta0} \\
g_{0}^{(1)} & = & 
\left[a_0\left(1+\frac{\kappa^2}{k_0^2}\right)\frac{\kappa^2}{m}\right]^{1/2},
\label{elementaryg0}
\\
\tilde{C}_0^{(1)} & = & -a_0\frac{\kappa^2}{k_0^2}.
\label{C0tilde}
\end{eqnarray}
In addition to the poles at $\pm i \kappa$,
$\kappa={\cal O}(\aleph)>0$,
the amplitude has zeros at $\pm k_0$, $|k_0|={\cal O}(\aleph)$.
The latter occurs on the real axis, $k_0>0$, 
for $0>\tilde{C}_0^{(1)}>-g_{0}^{(1)2}/|\Delta^{(0)}|$ 
--- this is the Ramsauer-Townsend effect \cite{Ramsauer:1921,Townsend:1922}. 
Otherwise, the zeros are on the imaginary axis. 
An example is the neutron-deuteron system in the spin-doublet 
channel \cite{Rupak:2018gnc}, although the bound state (the triton) 
and the virtual state are not symmetric with respect to the origin.
The sign of the scattering length $a_0={\cal O}(R)$ 
is constrained by the reality of $g_0^{(1)}$ \eqref{elementaryg0},
\begin{equation}
a_0\left(1+\frac{\kappa^2}{k_0^2}\right)>0.
\end{equation}

Although the amplitude \eqref{T0elementarydimer} has a pole on the positive
imaginary axis with a positive residue,
${\rm Res} (iS_0^{(1)}(k))|_{k=i\kappa} =\kappa a_0 (1+\kappa^2/k_0^2)>0$,
this is a pole of the potential itself. Just as in the diparticleless
formulation, the interpretation of this pole as an ``elementary'' particle
(in the EFT) seems justified.

\section{Resonances}
\label{narrow}

Much work in the recent literature 
has been dedicated to generalizing Weinberg's
criterion to resonances. A resonance is associated with two poles,
\begin{equation}
k_{\pm}
= \pm k_R- i k_I , 
\label{respoles}
\end{equation}
where, when they are shallow,
\begin{eqnarray}
k_I &=& \frac{1}{|r_{0}|} >0, 
\\
k_R &=& k_I \sqrt{\frac{2r_{0}}{a_{0}}-1} >0.
\label{resks}
\end{eqnarray}
Here, $2r_0<a_0<0$. For $2r_0<a_0< r_0$, the poles are closer
to the imaginary axis than to the real axis, $k_R< k_I$.
This is sometimes referred to as a subthreshold
resonance since the real part of its energy is negative. 
Its effect on scattering is limited
and the background phase shift is not negligible.
For $r_0< a_0<0$, in contrast, $k_R> k_I$: the proximity
of the resonance poles to the real axis makes the phase shift cross $\pi/2$
at the real (positive) part of the resonance energy.

From the EFT perspective, what stands out is that the ``elementary''
bound-state limit $Z\to 1$ shares the same power counting issues as 
narrow resonances \cite{Habashi:2020qgw}, 
for which $k_I\ll k_R$ as a consequence of $|r_0|\gg |a_0|$. 
This should not be surprising. 
Typically, a resonance emerges from a combination of attraction
and repulsion, and as attraction increases the two 
resonant poles collide on the negative imaginary axis at $-i/|r_0|$.
Narrow resonances will collide close to the origin, and as attraction 
increases further the bound state will emerge almost symmetrically with
respect to the virtual state. In other words, for large $|r_0|$
the relative splitting between the poles \eqref{kappasLO} is small,
reflecting the continuity with narrow resonances. 

Except in a very small window around it, a
narrow resonance can be described perturbatively with a diparticle field
\cite{Habashi:2020ofb}
just as a symmetric bound/virtual pair in the previous section.
Their shared property is a relatively large effective range $|r_0|\gg |a_0|$.
This suggests a simple generalization of Weinberg's parameter $Z$ \eqref{Z},
\begin{equation}
1-{\cal Z} \equiv \left(1+2\frac{|r_{0}|}{|a_{0}|}\right)^{-1/2},
\qquad 1>{\cal Z}>0.
\label{calZ}
\end{equation}
This quantity has in fact been proposed recently \cite{Matuschek:2020gqe}
as an extension to resonances that preserves the bounds on $Z$. 
An alternative quantity that has a simpler expression in terms of 
pole positions is
\begin{equation}
1-\tilde{\cal Z} \equiv 
\left|\mp\frac{i}{2k_\pm}
\left.\mathrm{Res}\!\left(iS_0^{(0)}\right)\right|_{k_\pm} \right|
= \left|\frac{k_+ + k_{-}}{k_{+} - k_{-}}\right|
= \left\{
\begin{array}{l}
1-Z \quad (\mathrm{bound/virtual}\; \mathrm{pair}),
\\
k_I/k_R \quad (\mathrm{resonance}),
\end{array}
\right.
\label{calZtilde}
\end{equation}
which falls in the ranges 
\begin{equation}
\left\{
\begin{array}{l}
0< 1-\tilde{\cal Z} <1  
\qquad (\mathrm{bound/virtual}\; \mathrm{pair} \; 
\mathrm{or} \; \mathrm{narrow} \; \mathrm{resonance}),
\\
1< 1-\tilde{\cal Z} <\infty \qquad (\mathrm{subthreshold}\; \mathrm{resonance}).
\end{array}
\right.
\end{equation}
This in turn corresponds to the proposal in Ref. \cite{Hyodo:2013iga}.

Both of these quantities account for the view of the diparticle as elementary
--- in the sense of weakly coupling and responsible for the poles ---
as ${\cal Z}\to 1$ or $\tilde{\cal Z}\to 1$
for either a narrow resonance or a bound/virtual pair.
In this limit the two symmetric
real poles collide at the origin
to produce two symmetric imaginary poles. 
For $|r_0|\gg |a_0|$, the two quantities agree,
$1-{\cal Z}\simeq \sqrt{|a_0|/2|r_0|} \simeq 1-\tilde{\cal Z}$.
Within $1-{\cal Z}={\cal O}(\aleph R)$
of this limit, 
the simplest formulation of the EFT consists of a 
bare diparticle propagator and a non-derivative contact interaction at LO, 
with subleading orders including loops and derivative interactions.

For $|a_0|\gg |r_0|$, both quantities are small, 
${\cal Z}\simeq |r_0|/|a_0|$ and $\tilde{\cal Z}\simeq |r_0|/a_0$.
Within the region ${\cal Z}= |\tilde{\cal Z}|={\cal O}(\aleph R)$, 
the low-energy dynamics is dominated by a single bound-state or virtual-state 
pole. However
${\cal Z}$ and $\tilde{\cal Z}$ differ for $a_0\sim r_0<0$ 
\cite{Matuschek:2020gqe}. 
The difference is particularly large for a subthreshold resonance 
but this type of resonance has limited impact on observables despite
being shallow.
From the EFT perspective, the main drawback of $\tilde{\cal Z}$ is
that it vanishes when $k_I=k_R$ without having an obvious association
with a change in power counting. In contrast, on that line of the
complex momentum plane ${\cal Z}=1-1/\sqrt{2}$.
Thus only ${\cal Z}$ tracks the changes in power counting
needed in the two special limits ${\cal Z}\to 0,1$.

\section{Conclusion}
\label{conc}

I have looked at Weinberg's ancient compositeness criterion through the glasses
of Weinberg's modern framework of effective field theories.
I employed two versions of an EFT for short-range interactions,
which incorporates the situation he originally considered of
two low-energy two-body $S$-wave poles: 
a bound state and a virtual state.
In the first version, only contact interactions are present.
In the second, an auxiliary diparticle field is included
to represent correlations among contact interactions.

In this EFT, 
Weinberg's parameter $Z$ \eqref{Z} is an observable related to the relative
pole distance in the relative momentum plane.
Its value can thus
be used as a criterion for {\it something}, but it is not immediately
clear at the level of poles why it is for compositeness.
Most of the literature focuses on quantities
that are scheme and/or model dependent. 
In contrast, EFT provides a model-independent
framework to describe observables. At least at the two-body level discussed
here, the two poles can, in most arrangements, be equivalently described
by either version of the EFT. This includes the limit $Z\to 0$
where the virtual pole is outside the regime of validity
of the EFT. The equivalence between the two versions of the EFT in
that case is well known (see {\it e.g.} Ref. \cite{vanKolck:1998bw}).
Neither the existence of two poles nor the convenience of
a diparticle field provide a compelling justification for Weinberg's
criterion from the EFT perspective.

It is only the $Z \to 1$ limit that 
leads to significant challenges to the diparticleless formulation.
It requires a perturbative nonlocal interaction that seems at odds with the
usual requirement of an expansion in derivatives at the Lagrangian
level. In the diparticle formulation, in the same limit 
the dominant interaction is still perturbative but energy dependent,
being produced by the $s$-channel exchange of the diparticle. 
The bound state arises from the diparticle propagator,
as one would expect when two particles couple weakly to
an ``elementary'' particle that shares the two-particle quantum numbers.
The unusual correlation among momentum-dependent interactions in 
the diparticleless formulation can be justified {\it a posteriori}
by the integrating-out of the ``elementary'' diparticle.
It is an example of Weinberg's {\it dictum}: ``You may use any degrees
of freedom you like to describe a physical system, but if you use the
wrong ones, you'll be sorry'' \cite{Weinberg:1981qq}.
This seems to me the strongest rationale for the notion of ``elementarity''.

There is a similarity between this limit and the extreme case of 
a zero-width resonance, both stemming from a relatively large
effective range, $|r_0|\gg |a_0|$. This similarity suggests the generalization
${\cal Z}$ \eqref{calZ} of Weinberg's $Z$, which was
proposed recently in Ref. \cite{Matuschek:2020gqe}
on the basis of a different but related argument.
The limit ${\cal Z}\to 1$ can be considered
that of an ``elementary'' diparticle, in the sense described above.
If the diparticle mass is above the two-particle threshold it generates
two narrow resonance poles, and if below, a bound/virtual pair.
Why the diparticle mass is close enough to the two-particle threshold for
the two poles to be shallow is not a question one can in general address 
without more information about the underlying theory.

Although this criterion provides a reasonable interpretation
of ``elementarity'',
in the end it is just a classification of
different power countings for the EFT with underlying
physics at distance ${\cal O}(R)$ and two shallow poles 
at momenta ${\cal O}(\aleph)$:
\begin{itemize}
\item For ${\cal Z}={\cal O}(\aleph R)$, physics is dominated by a single pole,
      which represents a bound or virtual state.
      The kinetic term of the diparticle field is an NLO effect. 
      If the diparticle is integrated out,
      the only LO interaction is a non-derivative contact interaction.
\item For ${\cal Z}\sim 1/2$, the two poles are important.
      The kinetic term of the diparticle field is LO. 
      If the diparticle is integrated out,
      the LO interactions consist of non-derivative and two-derivative
      contact interactions.
\item For $1- {\cal Z}={\cal O}(\aleph R)$, the two poles are nearly
      symmetrical with respect to the origin of the complex momentum 
      plane, either on the imaginary axis or very close to the real axis. 
      The poles arise from a weakly coupled diparticle or, if it is 
      integrated out, from an infinite number of 
      correlated contact interactions.
\end{itemize}
Different power countings are, ultimately,
a reflection of the size of observable quantities like $a_0$ and $r_0$.
The impact of these different power countings on more-body
systems should be studied \cite{Griesshammer:2023}.

Useful as it is for an EFT of short-range forces, which is
implicit in Weinberg's original paper \cite{Weinberg:1965zz},
${\cal Z}$ is in principle limited to this EFT.
In applications to exotic hadrons, where experimental
information is limited, the first issue to sort out is what
EFT to use. For example, for the $X(3872)$, an EFT with (perturbative)
pions \cite{Fleming:2007rp,Braaten:2015tga}
could replace a short-range EFT \cite{Braaten:2003he}
if the pole momenta are not
small compared to the inverse force range 
$R^{-1}\sim \sqrt{\Delta^2-m_\pi^2}$, where $m_\pi$ 
is the neutral-pion mass and $\Delta$ is the mass splitting 
between the charmed mesons $D^{*0}$ and $D^0$.
What role ${\cal Z}$ can play in the presence of 
such long-range forces and/or additional thresholds is a matter for further 
investigation.
One can mine for gold even in Steve's less well-known work!

\section*{Acknowledgments}
I thank Sean Fleming and Harald Grie{\ss}hammer for useful discussions
and comments on the manuscript.
I am grateful to the Kavli Institute for Theoretical Physics 
for hospitality while some of this work was carried out. 
This material is based upon work supported in part 
by the U.S. Department of Energy, Office of Science, Office of Nuclear Physics, 
under award DE-FG02-04ER41338
and by the National Science Foundation under Grant No. NSF PHY-1748958.

\end{document}